\documentclass[12pt]{article}

\def\lromn#1{\uppercase\expandafter{\romannumeral#1}}

\def\blist{\begin{list}{\setlength{\rightmargin}{\leftmargin}}}
\def\elist{\end{list}}
\begin{document}

\begin{flushright}
TU/00/605 
\end{flushright}

\begin{center}
\begin{large}

\renewcommand{\thefootnote}{\fnsymbol{footnote}}
\textbf{
Tunneling Phenomena in Cosmology:\\
Some Fundamental Problems
}
\footnote[1]
{
Talk given at COSMO2K, held at Cheju, Korea, 
September 4-8, 2000.
To appear in the Proceedings (World Scientific).
}

\end{large}

\vspace{4cm}
\begin{large}
M. Yoshimura

Department of Physics, Tohoku University\\
Sendai 980-8578 Japan
\end{large}

\vspace{4cm}

{\bf ABSTRACT}

\end{center}

The fundamental problem of how tunneling in thermal medium
is completed is addressed, and a new time scale of order
1/friction for its termination, 
which is usually much shorter than the Hubble time, is pointed out.
Enhanced non-linear resonance is responsible for this short time scale.
This phenomenon occurs when the semiclassical periodic
motion in a metastable potential well resonates with one of the
environment harmonic oscillators coupled to its motion.

\newpage

The usual picture of how the first order phase transition proceeds
in cosmology,
for instance the electroweak phase transition which is relevant for
the scenario of electroweak baryogenesis \cite{ew-bgeneration-review},
is something like this;
nucleation of the bubble of the true ground state (of broken symmetry)
occurs with
some probability in different parts of the universe.
If the bubble formation does not take place strongly,
the phase transition is terminated only when cosmological
evolution changes the form of the potential (or more properly
of the free energy) so that the symmetric phase is no
longer the local minimum of the potential.
This picture relies on the presumption that there is no
inherent time scale for termination of the phase transition,
or at least even if it exists, it is much larger than the
Hubble time scale, which is the time for the potential change.
We would like to exmaine this problem by working in a new
formalism of the real-time description of the tunneling 
phenomena \cite{my 00-2}, \cite{my 00-1}.
Clearly, if the new time scale of the phase transition is
much shorter than the Hubble time, one must reconsider the usual
scenario. This might change the situation on, and even 
resurect the once failed
GUT phase transition for the inflationary universe scenario
\cite{old inflation}.

Another important ingredient for our consideration
is effect of cosmic environment;
the order parameter that describes the state of the universe
such as the homogeneous part of the Higgs field
is necessarily coupled to matter fields that make up the thermal environment.
One must then take into accout the presence of thermal medium
in discussion of the tunneling rate and its time
evolution, and
estimate how dissipation due to the environment interaction
modifies the basic tunneling rate.

There are already many works on the subject
of tunneling in medium
\cite{q-tunneling review1}, and
most past works deal with a system in equilibrium as a whole.
The Euclidean technique such the bounce solution
\cite{bounce} is often
used in this context \cite{caldeira-leggett 83},
\cite{qbm review}.
Our approach here is different, and we attempt
to clarify dynamics of time evolution starting from
an arbitrary initial state of the tunneling system,
which can be either a pure or a mixed state.
We find it more illuminating to use a real-time
formalism instead of the Euclidean method much 
employed in the literature. 

Our formalism is based on separation of a subsystem from thermal
environment, and integrating out the environment variable
including the interaction with the subsystem.
In this picture dissipation seen in the behavior of the subsystem
is due to our ignorance of
huge environment degrees of freedom.
Modeling of the environment and its interaction form with
the subsystem is expected to be insensitive to the result one
derives in this approach.
We use the standard model of environment \cite{feynman-vernon},
\cite{caldeira-leggett 83}.

This method is  best suited to a (by itself)
non-equilibrium system which is immersed in a larger thermal equilibrium
state.
This methodology has been used by us in a number of problems
in cosmology \cite{decay jmy}, \cite{relic abundance my}.
For instance, the relic abundance was calculated \cite{relic abundance my} 
from this approach and
the usual estimate of thermally averaged Boltzmann rate
was justified at high temperatures.
At very low temperatures $T \ll M$ (the mass of relic particle)
our calculation differs from the usual one;
the Boltzmann suppressed number density $(MT/2\pi )^{3/2}\,
e^{-\,M/T}$ at the freeze-out
is replaced by some temperature power term.
(There is some criticism against this calculation
\cite{relic abundance criticism}.)
But numerical importance of this effect is presumably minor,
although it is a theoretically interesting issue.

Here we consider the most basic problem of this kind,
one dimensional subsystem described by a potential
$V(q)$.
This system is put in thermal medium.
The potential is assumed to have some local minimum at
$q = 0$, which is separated at $q = q_{B}$ of the barrier top
from a global minimum.
The environment part is modeled by infinitely many, continuously
distributed harmonic oscillators whose coordinates are
$Q(\omega )$.
Its coupling to the tunneling system is given by a Hamiltonian,
\begin{equation}
q\,\int_{\omega _{c}}^{\infty }\,d\omega \,c(\omega )Q(\omega )
\,.
\end{equation}
The coupling strength is $c(\omega )$ and $\omega _{c}$ is some
threshold frequency.
Needless to say, one may imagine a generalized case
in which the system variable
$q$ is the order parameter for the first order phase transition,
the homogeneous Higgs field, and the environment oscillator
$Q(\omega )$ is a collection of various forms of matter fields
coupled to the Higgs field.

The basic equation in our problem is
\begin{eqnarray}
&&
\frac{d^{2}q}{dt^{2}} + \frac{dV}{dq} =
-\,\int_{\omega_c}^{\infty}\,d\omega \,c(\omega )Q(\omega ) \,, 
\hspace{0.5cm}
\frac{d^{2}Q(\omega )}{dt^{2}} + \omega ^{2}\,Q(\omega )
= -\,c(\omega )\,q
 \,.
\label{eq of motion}
\end{eqnarray}
One may eliminate the environment variable $Q(\omega )$  to get
the Langevin equation \cite{ford-lewis-oconnell},
\begin{eqnarray}
&&
\frac{d^{2}q}{dt^{2}} + \frac{d V}{d q} +
2\,\int_{0}^{t}\,ds\,\alpha _{I}(t - s)q(s) = F_{Q}(t) 
\,,
\label{langevin eq}
\end{eqnarray}
where $F_{Q}(t) $ is linear in initial values of environment variables,
$Q_{i}(\omega )$ and $P_{i}(\omega)$;
\begin{eqnarray}
&&
F_{Q}(t) = -\,
\int_{\omega _{c}}^{\infty }\,d\omega \,c(\omega )\,
\left( \,Q_{i}(\omega ) \cos (\omega t) +
\frac{P_{i}(\omega)}{\omega } \sin (\omega t)\,\right)
\,.
\end{eqnarray}
By taking the thermal bath of temperature $T = 1/\beta $
given by the density matrix,
\begin{eqnarray}
\rho _{\beta }(Q \,, Q') &=& \left( \frac{\omega }{\pi \,\coth (\beta \omega 
/2)}\right)^{1/2}\,
\nonumber \\ &&
\cdot \exp \left[ \,-\,\frac{\omega }{2 \sinh (\beta \omega )}
\,\left( \,(Q^{2} + Q'\,^{2})\,\cosh (\beta \omega ) - 2 Q Q'\,\right)\,\right]
\,,
\label{thermal density matrix} 
\end{eqnarray}
for each environment oscillator,
the following correlation formula is obtained;
\begin{eqnarray}
&&
\hspace*{-0.5cm}
\langle \{ F_{Q}(\tau )\,,F_{Q}(s) \}_{+} \rangle_{{\rm env}}
= \int_{\omega _{c}}^{\infty }\,d\omega \,r(\omega )
\cos \omega (\tau - s)\,\coth (\frac{\beta \omega }{2}) 
\equiv \alpha _{R}(\tau - s)
\,,
\end{eqnarray}
with 
\begin{equation}
r(\omega ) = \frac{c^{2}(\omega )}{2\omega}
\,.
\end{equation}
The kernel function $\alpha _{I}$ in eq.(\ref{langevin eq}) 
is given by
\begin{equation}
\alpha _{I}(t) = -\,\int_{\omega _{c}}^{\infty }\,d\omega \,
r(\omega )\sin (\omega t) 
 \,.
\end{equation}
The combination, $\alpha _{R}(t) + i\,\alpha _{I}(t)$,
is a sum of the real-time thermal Green's function for
$Q(\omega)-$oscillators, added with
the weight $c^2(\omega)$.

An often used simplification is the local, Ohmic approximation
taking 
\begin{equation}
r(\omega) = \frac{\eta\,\omega}{\pi} \,,
\end{equation}
with $\omega_c = 0$,
which amounts to
\begin{equation}
\alpha _{I}(\tau ) = \delta \omega ^{2}\delta (\tau )
+ \eta\, \delta '(\tau ) \,.
\end{equation}
This gives the local version of Langevin equation,
\begin{equation}
\frac{d^{2}q}{dt^{2}} + \frac{d V}{d q} + 
\delta \omega ^{2}\,q + \eta\,\frac{dq}{dt} = 0 \,.
\end{equation}
The parameter $\delta \omega ^{2}$ is interpreted as
a potential renormalization or a mass renormalization in the
field theory analogy, since by changing the bare frequency parameter
to the renormalized $\omega _{R}^{2}$ the term
$\delta \omega ^{2}\,q$ is cancelled by a counter term
in the potential.
On the other hand, $\eta $ is the Ohmic friction coefficient.
This local approximation breaks down both at early and at late
times \cite{difficulty of ohm}, 
but it is useful in many other cases.

Any mixture of quantum states is described by a density matrix,
\begin{equation}
\rho = \sum_{n}\,w_{n}\,|n\rangle \langle n |\,, 
\label{def density matrix} 
\end{equation}
where $|n\rangle$ is a state vector for pure quantum state and
$0 \leq w_{n} \leq 1$.
The density matrix obeys the equation of motion;
\begin{equation}
i\hbar \frac{\partial \rho }{\partial t} =
\left[ \,H _{{\rm tot}} \,, \: \rho \,\right]
\,, 
\label{liouville eq} 
\end{equation}
where $H _{{\rm tot}}$ is the total Hamiltonina of the entire
system.
In configuration space this density matrix is given by its
matrix elements,
\begin{equation}
\rho (q \,, q' \,; \; Q(\omega ) \,, Q'(\omega ) )
= \langle q \,, Q(\omega )|\,\rho \,|q' \,, Q'(\omega ) \rangle
\,.
\end{equation}
Its Fourier transform with respect to relative coordinates,
\( \:
q - q' \,, Q(\omega ) - Q'(\omega )
\,, 
\: \)
is called the Wigner function and denoted by $f_{W}$;
\begin{eqnarray}
&&
\hspace*{1cm} 
f_{W}(q \,, p \,; \; Q(\omega )\,, P(\omega ) )
= \int_{-\infty }^{\infty }\,d\xi \,\prod\,dX(\omega )
\nonumber \\ &&
\hspace*{-1cm}
\exp [-\,i\xi p - i\int\,d\omega \,X(\omega )P(\omega )]\,
\rho (q + \xi /2\,, q - \xi /2 \,; \; Q(\omega ) +
X(\omega )/2\,, Q(\omega ) - X(\omega )/2 ) \,.
\nonumber \\ &&
\end{eqnarray}
It is easy to show from eq.(\ref{liouville eq}) that this quantity obeys
\begin{eqnarray}
&&
\hspace*{0.5cm} 
\frac{\partial f_{W}}{\partial t} =
-\,p\,\frac{\partial f_{W}}{\partial q}
- \int\,d\omega \,
\left( \,
P(\omega )\,\frac{\partial f_{W}}{\partial Q(\omega )}
+ \omega ^{2}\,Q(\omega )\,
\frac{\partial f_{W}}{\partial P(\omega )}
\right.
\nonumber \\ 
\hspace*{-0.5cm} 
&&
\left.
+ c(\omega )\,(q\frac{\partial }{\partial P(\omega )}
+ Q(\omega )\frac{\partial }{\partial p})\,f_{W}
\,\right)
+ \frac{1}{i\hbar }\,
\left\{ \,
V\left( q + \frac{i\hbar }{2}\frac{\partial }{\partial p}\right) 
- V\left( q - \frac{i\hbar }{2}\frac{\partial }{\partial p}\right) 
\,\right\}
f_{W}
\,. 
\nonumber \\ &&
\label{master eq for wigner} 
\end{eqnarray}

In one dimensional quantum mechanics the semiclassical approximation
is excellent when the potential barrier is large, and
we assume that this is also true in the presence of the 
system-environment interaction.
In the semiclassical $\hbar \rightarrow 0$ limit we have
\begin{eqnarray}
\frac{1}{i\hbar }\,
\left\{ \,
V\left( q + \frac{i\hbar }{2}\frac{\partial }{\partial p}\right) 
- V\left( q - \frac{i\hbar }{2}\frac{\partial }{\partial p}\right) 
\,\right\}
f_{W}
\,\rightarrow\,
\frac{d V}{d q}\,\frac{\partial f_{W}}{\partial p}
\,.
\end{eqnarray}
The resulting equation, being identical to the classical Liouville equation,
has an obvious solution;
\begin{eqnarray}
&&
f_{W}(q\,, p\,, Q\,, P) = 
\int\,dq_{i}dp_{i}\,\int\,dQ_{i}dP_{i}\,
f_{W}^{(i)}(q_{i}\,, p_{i}\,, Q_{i}\,, P_{i})\,
\nonumber \\ &&
\hspace*{1cm} 
\cdot 
\delta \left( q - q_{{\rm cl}}\right)\,\delta \left( p - p_{{\rm cl}}\right)
\,\delta \left( Q - Q_{{\rm cl}}\right)\,\delta \left( P 
- P_{{\rm cl}}\right)
 \,,
\end{eqnarray}
where 
\( \:
q_{{\rm cl}}(q_{i}\,, p_{i}\,, Q_{i}\,, P_{i}\, ; t)
\: \)
etc. are the solution of 
(\ref{eq of motion}), 
taken as the classical equation with the specified initial condition.

We consider the circumstance under which the tunneling system
is initially in a state uncorrelated to the rest of environment.
Thus we take the form of the density matrix,
\( \:
\rho ^{(i)} = \rho _{q}^{(i)} \otimes \rho _{Q}^{(i)}
 \,,
\: \)
to get the reduced Wigner function after the trivial
$Q(\omega) \,, P(\omega)$ integration,
\begin{eqnarray}
&&
f_{W}^{(R)}(q \,, p\,;t) = 
\int\,dq_{i}dp_{i}\,
f_{W\,, q}^{(i)}(q_{i}\,, p_{i})\,K(q \,, p\,, q_{i}\,, p_{i}\,;t)
\,,
\label{integral transform} 
\\ && 
K(q \,, p\,, q_{i}\,, p_{i}\,;t) = \int\,dQ_{i}dP_{i}\,
f_{W\,, Q}^{(i)}(Q_{i}\,, P_{i})\,
\delta \left( q - q_{{\rm cl}}\right)\,\delta \left( p - p_{{\rm cl}}\right)
 \,.
\end{eqnarray}
The problem of great interest is how further one can simplify
the kernel function $K$ here.

In many situations one is interested in the tunneling probability when
the environment temperature is low enough.
At low temperatures of
\( \:
T \ll 
\: \)
a typical frequency or curvature scale 
$\omega _{s}$ of the potential $V$,
one has
\begin{equation}
\omega _{s}\,\sqrt{\,\overline{Q_{i}^{2}(\omega )}\,} \,, 
\sqrt{\,\overline{P_{i}^{2}(\omega )}\,} = O[\sqrt{T}]
\ll \sqrt{\omega _{s}}
 \,.
\end{equation}
In the electroweak and the GUT phase transition this
corresponds to $T \ll m_{H}$(Higgs mass).
Expansion of $q_{{\rm cl}}$ in terms of 
\( \:
Q_{i}(\omega ) \,, P_{i}(\omega )
\: \)
is then justified.
Thus, we use
\begin{eqnarray}
&&
\delta \left( \,q - q_{{\rm cl}}\,\right)
= \int\,\frac{d\lambda _{q}}{2\pi }\,
\exp \left[ \,i\,\lambda _{q}\,\left( \,q - q_{{\rm cl}}\,\right)\,\right]
\nonumber \\ &&
\hspace*{-1cm}
\approx \int\,\frac{d\lambda _{q}}{2\pi }\,
\exp \left[ \,i\,\lambda _{q}\,
\left( \,q - q_{{\rm cl}}^{(0)} -
\int\,d\omega \,
\left\{\,Q_{i}(\omega )q_{{\rm cl}}^{(Q)}(\omega )
+ P_{i}(\omega )\,q_{{\rm cl}}^{(P)}(\omega )\,
\right\}\,\right)\,\right]
 \,,
\label{expansion of cl sol}
\end{eqnarray}
valid to the first order of $Q_{i}(\omega) \,, P_{i}(\omega)$.
A similar expansion for
\( \:
\delta \left( \,p - p_{{\rm cl}}\,\right)
\: \)
using
\( \:
p_{{\rm cl}}^{(0)} \,, p_{{\rm cl}}^{(Q)}(\omega ) \,, 
p_{{\rm cl}}^{(P)}(\omega )
 \,,
\: \)
also holds.

An alternative justification of the expansion (\ref{expansion of cl sol})
is to keep $O[c(\omega)]$ terms consistently in the exponent
since both $q_{{\rm cl}}^{(Q)}(\omega )$ and $q_{{\rm cl}}^{(P)}(\omega )$
are of this order.

Gaussian integral for the variables
\( \:
Q_{i}(\omega ) \,, P_{i}(\omega ) \,
\: \)
first and then for
\( \:
 \lambda _{q} \,, \lambda _{p}
\: \)
can be done explicitly with eq.(\ref{expansion of cl sol}), 
when one takes the thermal, hence Gaussian density matrix
for the initial environment variable, $\rho _{Q}^{(i)}$.
The result of this Gaussian integral leads to
an integral transform \cite{my 00-2} of the Wigner function, 
$f_{W}^{(i)} $ (initial) $\rightarrow f_{W}^{(R)}$ (final), 
using the kernel function of
\begin{eqnarray}
&&
\hspace*{-1cm}
K(q \,, p\,, q_{i}\,, p_{i}\,;t) = 
\frac{\sqrt{{\rm det}\; {\cal J}\,}}{2\pi}
\exp \left[ \,-\,\frac{1}{2}
(q - q_{{\rm cl}}^{(0)}\,, \, p - p_{{\rm cl}}^{(0)})
\,{\cal J}\,
\left( \begin{array}{c}
q - q_{{\rm cl}}^{(0)}  \\
p - p_{{\rm cl}}^{(0)}
\end{array}
\right)
\,\right]
 \,,
 \label{kernel function} 
\end{eqnarray}
where the matrix elements of 
\( \:
{\cal J}^{-1} = I_{ij}
\: \)
are given by 
\begin{eqnarray}
&&
I_{11} =\frac{1}{2}\,\int_{\omega _{c}}^{\infty }\,d\omega \,
\coth \frac{\beta \omega }{2}\,
\frac{1}{\omega } \,|z(\omega \,, t)|^2
\,, 
\label{fluctuation 11}
\\ &&
I_{22} =\frac{1}{2}\,\int_{\omega _{c}}^{\infty }\,d\omega \,
\coth \frac{\beta \omega }{2}\,
\frac{1}{\omega } \,
|\dot{z}(\omega \,, t)|^2
\,, 
\\ &&
\hspace*{-1cm}
I_{12}  =\frac{1}{2}\,\int_{\omega _{c}}^{\infty }\,d\omega \,
\coth \frac{\beta \omega }{2}\,
\frac{1}{\omega } \,\Re\left( z(\omega \,, t)\dot{z}^*(\omega \,, t)\right)
= \frac{\dot{I}_{11}}{2\,I_{11}}
\,,
\end{eqnarray}
where
\( \:
z(\omega \,, t) = 
q_{{\rm cl}}^{(Q)}(\omega \,, t) + i\,
\omega \,q_{{\rm cl}}^{(P)}(\omega \,, t)\,,
\: \)
and 
\( \:
\dot{z}(\omega \,, t) = 
p_{{\rm cl}}^{(Q)}(\omega \,, t) +
i\omega \,p_{{\rm cl}}^{(P)}(\omega \,, t)
\,.
\: \)

Quantities that appear in the integral transform are determined by solving
differential equations; 
the homogeneous Langevin equation for $q_{{\rm cl}}^{(0)}$
and an inhomogeneous linear equation for $z(\omega \,, t)$ and
$\dot{z}(\omega \,, t)$,
\begin{eqnarray}
&&
\frac{d^{2}q_{{\rm cl}}^{(0)}}{dt^{2}} +
\left( \frac{d V}{dq }\right)_{q_{{\rm cl}}^{(0)}}
+ 2\,\int_{0}^{t}\,ds\,\alpha _{I}(t - s)q_{{\rm cl}}^{(0)}(s) = 0
 \,,
\label{homo classical eq}
\\ &&
\frac{d^{2}z(\omega \,, t)}{dt^{2}} +
\left( \frac{d^{2} V}{dq^{2} }\right)_{q_{{\rm cl}}^{(0)}}\,z(\omega \,, t)
+ 2\,\int_{0}^{t}\,ds\,\alpha _{I}(t - s)z(\omega \,, s) = 
-\,c(\omega )e^{i\omega t} \,.
\label{fluctuation eq}
\end{eqnarray}
A similar equation as for $z(\omega \,, t)$ holds for
\( \:
\dot{z}(\omega \,, t)  \,.
\: \)
The initial condition is \\
\( \:
q_{{\rm cl}}^{(0)}(t = 0) = q_{i} \,, \hspace{0.5cm} 
p_{{\rm cl}}^{(0)}(t = 0) = p_{i} \,, \hspace{0.5cm} 
z(\omega \,, t = 0) = 0  \,, \hspace{0.5cm}
\dot{z}(\omega \,, t = 0) = 0 \,.
\: \)

The physical picture underlying the formula for the integral
transform, eq.(\ref{integral transform}) along with 
(\ref{kernel function}), should be evident;
the probability at a phase space point $(q\,,p)$ is dominated 
by the semiclassical trajectory $q_{{\rm cl}}^{(0)}$
(environment effect of dissipation being included in its determination
by eq.(\ref{homo classical eq}))
reaching $(q\,,p)$ from an initial point $(q_i \,, p_i)$ 
whose contribution is weighed by the quantum mechanical probability 
$f^{(i)}_W$ initially given.
The contributing trajectory is broadened by the environment
interaction with the width factor $\sqrt{I_{ij}}$.
The quantity $I_{11}$ given by (\ref{fluctuation 11}),
for instance, is equal to
\( \:
\overline{(q - q_{{\rm cl}}^{(0)})^{2}}
\,;
\: \)
an environment driven fluctuation under the stochastic
force $F_{Q}(t)$.

We note that in the exactly solvable model of inverted
harmonic oscillator potential the identical form of
the integral transform was derived \cite{my 00-1} without resort to
the semiclassical approximation.
Explicit form of the classical and the fluctuation
functions ($q_{{\rm cl}}^{(0)}$ and $z(\omega \,, t)$) is given there.

To proceed further, 
one may separate the tunneling potential $V(q)$ at the barrier
top location, $q = q_{B}$.
We distinguish two cases of the potential type,
depending on the value of $V(\infty)$ relative to the local minimum value 
$V_m$ at $q = 0$ in the potential well.
When $V(\infty) < V_m$, the classical motion in
the overbarrier region $q > q_{B}$ is monotonic
ending at $q = \infty$,
while for $V(\infty) > V_m$  the motion is a damped oscillation
towards $q_0$ at the true minimum, unless the friction is large.
If the friction is larger than a critical value of
$\approx 2\omega_*$ with $\omega_*$ being the curvature of the
potential at the global minimum, there occurs the overdamping 
such that $q_{\rm{cl}}^{(0)} \rightarrow q_0$ monotonically.
A typical interesting case for $V(\infty) > V_m$
is the asymmetric double well as may occur in the first order
electroweak phase transition \cite{ew-bgeneration-review}.

We consider here a few problems
to illustrate consequences of our general formula of
the integral transform.
One problem is calculation of the barrier penetration factor;
to determine the outgoing flux in the overbarrier region
for the type of potential of $V(\infty) < V_m$, assuming an initial 
energy eigenstate under the potential $V$. 
The other problem is the tunneling probability for the asymmetric
wine bottle type of potential.  

Energy eigenstates are special among pure quantum states,
since they evolve only with phase factors.
Thus, if one takes for $|n \rangle \langle n |$ in 
eq.(\ref{def density matrix})
the energy eigenstate of the total Hamiltonian,
the density matrix does not change with time.
On the other hand, if one takes the energy eigenstate
for the subsystem Hamiltonian, then the density matrix
changes solely due to the environment interaction.
It is thus best to use a pure eigenstate, or its superposition, of
the subsystem Hamiltonian when one wishes to determine how the barrier
penetration factor is modified in thermal medium.

We first discuss the stationary phase approximation for the
initial state.
Starting from the formula,
\begin{eqnarray}
&&
f_{W}^{(i)}(q_{i} \,, p_{i}) = \int\,d\xi \,e^{-\,ip_{i}\xi }\,
\sum_{n}\,w_{n}\,
\psi _{n}^{*}(q_{i} - \frac{\xi }{2})\psi _{n}(q_{i} + \frac{\xi }{2})
= 
\nonumber \\ &&
\hspace*{0.5cm} 
\sum_{n}\,w_{n}\,\int\,d\xi \,
\exp \left[ \,-\,ip_{i}\xi + \ln \psi _{n}^{*}(q_{i} - \frac{\xi }{2})
+ \ln \psi _{n}(q_{i} + \frac{\xi }{2})
\right]  \,,
\end{eqnarray}
we locate the stationary point by
\( \:
\frac{\partial }{\partial \xi } = 0 \,, 
\frac{\partial }{\partial p_{i}} = 0
\: \)
for each exponent factor. 
The partial derivative $\frac{\partial }{\partial p_{i}} = 0$
is taken with the understanding that the rest of $p_{i}$ integration
contains smooth functions of $p_{i}$.
This leads to the stationary point at
\begin{eqnarray}
&&
\xi = 0 \,, \hspace{0.5cm} 
p_{i} = I_{n}(q_{i}) 
\\ &&
I_{n}(x) =
\frac{-\,i}{2}\,\frac{\psi_{n}^{*} (x)\dot{\psi }_{n}(x) -
\dot{\psi }_{n}^{*}(x)\psi_{n} (x)}{|\psi_{n} (x)|^{2}}
 \,.
\end{eqnarray}
Here $I_{n}(x)$ is the usual flux factor at $x$ for
the pure state $|n\rangle$.
This gives the result for the initial density matrix,
\begin{equation}
f_{W}^{(i)}(q_{i} \,, p_{i}) \approx 
\sum_{n}^{}\,w_{n}\,|\psi _{n}(q_{i})|^{2}\,
2\pi \,\delta \left( p_{i} - I_{n}(q_{i}) \right)
\,.
\end{equation}

The barrier penetration factor is calculated from 
the flux formula,
\begin{eqnarray}
&&
I(q\,, t) = \int_{-\infty }^{\infty }\,\frac{dp}{2\pi}\,
p\,f_{W}^{(R)}(q\,, p\,; t)
 \,.
\end{eqnarray}
Thus, the semiclassical plus the stationary phase approximation gives
\begin{eqnarray}
&& 
\hspace*{-1cm}
I(q\,, t) \approx 
\int_{q_{*}}^{\infty }\,dq_{i}\,|\psi (q_{i})|^{2}\,
\sqrt{\frac{1}{2\pi I_{11}}}\,\exp \left[ \,-\,
\frac{(q - q_{{\rm cl}}^{(0)})^{2}}{2I_{11}}\,\right]
\cdot \left( \,p_{{\rm cl}}^{(0)} + \frac{\dot{I}_{11}}{2I_{11}}\,
(q - q_{{\rm cl}}^{(0)})\,\right)
 \,,
\nonumber \\ &&
\end{eqnarray}
for a pure initial state given by the wave function $\psi (x)$, where
$q_{*}$ is the turning point in the overbarrier region.

In the region of $ q \gg q_*$ there is always a classical trajectory that 
reaches the point $q$ of the Gaussian peak from 
an initial $q_{i}$ in the range 
$q_i > q_{*}$, and the entire region within the Gaussian width 
$\sqrt{I_{11}}$ is fully covered by the integral.
The factor outside the Gaussian exponent is well approximated
by its peak value in the weak coupling case.
Thus, one obtains \cite{my 00-2}
\begin{eqnarray}
&&
I(q\,, t) \approx |\psi (x_{0})|^{2}\,\left( -\,\dot{x}_{0}\right)
= |\psi (x_{0})|^{2}\,p_{{\rm cl}}^{(0)}(x_{0} \,, t)
\left( \frac{dq_{{\rm cl}}^{(0)}(x_{0}\,, t)}{dx_{0}}\right)^{-1}
 \,,
\end{eqnarray}
where $x_{0}(q\,, t)$ is determined from
\begin{equation}
q_{{\rm cl}}^{(0)}(q_i = x_{0} \,, p_i = I(x_{0}) \,, 
Q_i(\omega) = 0 \,, P_i(\omega) = 0 \,; t)
= q \,.
\end{equation}

We now take the WKB wave function in the overbarrier region,
\begin{eqnarray}
&&
\psi (q_{i}) = \frac{T(E)}{\sqrt{p(q_{i})}}\,
\exp \left[ i\,\int_{x_{*}}^{q_{i}}\,dx\,p(x)\right]
 \,, \hspace{0.5cm} 
p(x) = \sqrt{\,2(E - V(x))\,}
\,, 
\end{eqnarray}
to get a factorized form of the flux \cite{my 00-2},
\begin{eqnarray}
&&
I(q\,, t) \approx |T(E)|^{2}\,f(q\,, t\,; E) \,, 
\hspace{0.5cm} 
f = \frac{p_{{\rm cl}}^{(0)}(x_{0} \,, t)}{p(x_{0})}
\left( \frac{dq_{{\rm cl}}^{(0)}(x_{0} \,, t)}{dx_{0}}
\right)^{-1}
 \,.
\end{eqnarray}
A feature that characterizes the classical trajectory $q_{{\rm cl}}^{(0)}$
is of special interest;
its initial energy is
\begin{equation}
H_{q}(t = 0) \approx \frac{1}{2}\, p(q_{i})^{2} +
V(q_{i}) = E
\,.
\end{equation}

This formula for the flux reproduces our previous result for a specific
potential of the inverted harmonic oscillator \cite{my 00-1},
\( \:
V(x) + \frac{1}{2}\, \delta \omega ^{2}\,x^{2}
= -\frac{1}{2}\,\omega_R ^2\,x^2 \,,
\: \)
with $\omega _{R}$ the renormalized curvature.
The present approach actually improves our previous result;
\begin{eqnarray}
&&
f = \frac{\stackrel{..}{g}x_{0} + \dot{g}I(x_0)}
{\dot{g}I(x_0) + g\omega _{B}^{2}x_{0}}
\rightarrow \frac{\stackrel{..}{g} + \omega _{B}\dot{g}}
{\omega _{B}(\dot{g} + \omega _{B}g)}
 \,,
 \label{modified flux for ivho}
\end{eqnarray}
where $\omega_B \approx \omega_R  - \eta/2$ is a diagonalized frequency.
The limiting formula of eq.(\ref{modified flux for ivho}), 
with $I(x_0) \rightarrow \omega _{B} x_{0}$ as
$x_{0} \rightarrow \infty $, is valid in the infinite $q$ limit,
as derived in \cite{my 00-1}.
The function $g(t)$ is the homogeneous solution of
the Langevin equation given explicitly in \cite{my 00-1};
\begin{eqnarray}
&&
g(t) = 
\frac{N^{2}}{\omega _{B}}\,\sinh (\omega _{B}t )
+ 2\,\int_{\omega _{c}}^{\infty }\,d\omega \,H(\omega )\sin (\omega t)
\\ &&
H(\omega ) \approx  \frac{r(\omega )}{(\omega ^{2} + \omega _{R}^{2} )^{2}
 + (\pi r(\omega ))^{2}}  \,,
\\ &&
N^{2} = 1 - 
2\,\int_{\omega _{c}}^{\infty }\,d\omega \,H(\omega )\omega 
\,.
\end{eqnarray}
Both at early and late times the factor $f \approx 1$,
deviating from unity only for the time range of order $1/\omega_B$.

According to the view of \cite{caldeira-leggett 83} 
the potential $V(q)$ that determines the semiclassical
penetration factor $|T(E)|^{2}$ should be expressed in terms
of the renormalized parameters, $\omega_R$ in this case.
This explains the bulk of the suppression caused by
dissipative environment interaction, as explained in
\cite{my 00-1}.

For discussion of a more general case of finite $V(\infty) < V_m$ 
we use the local, Ohmic approximation, 
which becomes excellent at late times.
A potential that decreases fast as $q \rightarrow \infty $ is assumed;
\( \:
\frac{dV}{dq} \rightarrow 0 \,.
\: \)
The acceleration term can then be neglected when the friction
satisfies 
$\eta ^{2} \gg |\frac{d^{2}V}{dq^{2}}| $.
This is a slow rolling approximation, and it always holds
for $q$ large enough.
The classical equation is then solved as
\begin{equation}
\eta \,\int_{q_{*}}^{q}\,dz(\frac{dV}{dz})^{-1} = - \,t
 \,,
\end{equation}
which gives the factor
\begin{equation}
f \approx 
-\,\frac{dV}{dq_{*}}\,\frac{1}{\eta p(\infty )}  \,.
\end{equation}
Thus, the tunneling probability decreases with time along with the decreasing
slope of the potential.
This result poses a curious question;
the tunneling may not be terminated within a limited finite
time. We shall encounter a similar situation for the asymmetric
wine bottle potential.

We next consider the case in which the potential is very steep
at both ends;
\( \:
V(\pm \infty ) = \infty \,.
\: \)
The tunneling rate, from the inner region at $q < q_B$
into the outer region at $q > q_B$, is an important measure
of tunneling phenomena and is given by the flux at $q = q_B$;
\( \:
\dot{P}(t) = -\,I(q_{B} \,, t) \,,
\: \)
which is equal to
\begin{eqnarray}
&&
\hspace*{-1cm}
-\,
\int\,dq_{i}dp_{i}\,f_{W}^{(i)}(q_{i} \,, p_{i})
\,\left(\, 
p_{{\rm cl}}^{(0)} + \frac{\dot{I}_{11}}{2\,I_{11}}\,
(q_{B} - q_{{\rm cl}}^{(0)}\,)
\,\right)
\frac{1}{\sqrt{2\pi I_{11}}}\,\exp \left[ \,-\,
\frac{(q_{B} - q_{{\rm cl}}^{(0)}\,)^{2}}{2I_{11}}\,\right]
 \,.
\nonumber \\ &&
\end{eqnarray}
On the other hand, the tunneling probability into the overbarrier
region at $q > x$ is given by
\begin{eqnarray}
&&
P(x \,, t) = 
\int\,dq_{i}dp_{i}\,f_{W}^{(i)}(q_{i} \,, p_{i})\,
\int_{x}^{\infty }\,du\,
\frac{1}{\sqrt{2\pi I_{11}}}\,\exp \left[ \,-\,
\frac{(u - q_{{\rm cl}}^{(0)}\,)^{2}}{2I_{11}}\,\right]
 \,.
\end{eqnarray}
In both of the quantities, $\dot{P}(t)$ and $P(q_B \,, t)$
(the tunneling probability into $q > q_B$), 
it is essential to
estimate how $q_{{\rm cl}}^{(0)}$ and $I_{11}$ varies with time.

We consider an initial state localized in the potential
well so that the dominant contribution in the $(q_{i} \,, p_{i})$
phase space integration is restricted to $q_{i} < q_{B}$.
In the rest of discussion anharmonic terms play important roles,
but at first we work out the harmonic approximation;
\begin{equation}
V(q) + \frac{1}{2}\,
\delta \omega^2\,q^2 \approx \frac{1}{2}\,\omega_0 ^2 \, q^2
\,, 
\end{equation}
near the bottom of the well at $q = 0$. 
In the Ohmic approximation the classical solution and
the fluctuation is given by
\begin{eqnarray}
&&
q_{{\rm cl}}^{(0)} = \left(\,\cos \tilde{\omega }_{0}t + \frac{\eta }{2
\tilde{\omega }_{0}}\sin \tilde{\omega }_{0}t \,\right)\,e^{-\eta t/2}\,q_{i}
 + \frac{\sin \tilde{\omega }_{0}t}{\tilde{\omega }_{0}}\,
e^{-\eta t/2}\,p_{i} \,, 
\\ && 
\hspace*{0.5cm} 
z(\omega \,, t) = 
\frac{c(\omega )}{\omega ^{2} - \omega _{0}^{2} - i\omega \eta }
\nonumber \\ &&
\cdot 
\left( \,
e^{i\omega t} - \frac{\omega + \tilde{\omega }_{0} - i\eta /2}
{2\tilde{\omega }_{0}}\,e^{i\tilde{\omega }_{0}t - \eta t/2}
+ \frac{\omega - \tilde{\omega }_{0} - i\eta /2}{2\tilde{\omega }_{0}}
\,e^{- i\tilde{\omega }_{0}t - \eta t/2}
\right)
 \,,
\end{eqnarray}
using
\( \:
\tilde{\omega }_{0} = \sqrt{\omega _{0}^{2} - \frac{\eta ^{2}}{4}}
\,.
\: \)
In the rest of discussion we assume a small friction,
\( \:
\eta \ll \omega _{0} \,.
\: \)

Near $\omega = \tilde{\omega }_{0}$ the fluctuation is approximately
\begin{eqnarray}
&&
z(\omega \,, t) \approx \frac{i\,c(\omega )}
{\omega + \tilde{\omega }_{0}- i\eta/2}
\,\left( \, t \,e^{i\tilde{\omega }_{0}t} - \frac{1}{\tilde{\omega }_{0}}\,
\sin \tilde{\omega }_{0}t\,\right)\,
e^{- \eta t/2}
 \,.
 \label{harmonic z}
\end{eqnarray}
This formula is valid at $t < O[1/\eta]$.
The appearance of the linear $t$ term is a resonance effect.
The resonance roughly contributes to $I_{11}(t)$ by the amount,
\( \:
\eta\, t^{2}\,  e^{-\eta t}\, \times
\, \)
a smooth $\omega $ integral which is cut off by a physical 
frequency scale.
Thus, the width factor $I_{11}$ 
initially increases with time until the time
scale of order $1/\eta$.

The width factor asymptotically behaves as
\begin{eqnarray}
&&
I_{11}(t) = I_{11}(\infty ) + O[e^{-\,\eta t/2}]
\,,
\\ &&
I_{11}(\infty ) \approx  \frac{1}{2\omega _{0}} + \frac{1}{\omega _{0}}\,
\frac{1}{e^{\omega _{0}/T} - 1} + \frac{\pi }{3}
\frac{\eta }{\omega _{0}^{4}}\,T^{2}  \,.
\end{eqnarray}
The asymptotic value of $I_{11}(\infty )$ has
the familiar zero point fluctuation of harmonic oscillator 
and in the last term the dominant finite temperature correction,
valid for this Ohmic model at $T \ll \omega_0$.
In any event the probability rate $\dot{P}(t)$ finally decreases 
to zero, along with
\begin{equation}
p_{{\rm cl}}^{(0)} + \frac{\dot{I}_{11}}{2I_{11}}\,
(q_{B} - q_{{\rm cl}}^{(0)}\,) \rightarrow 0
\,. 
\end{equation}
Moreover, the final tunneling probability $P(q_{B} \,, \infty )$
has a finite value, and typically is very small for a large
potential barrier.
For instance, for the asymmetric wine bottle potential shortly
discussed,
\begin{equation}
P(q_{B} \,, \infty ) \approx \frac{1}{4}\,
\sqrt{\,\frac{\omega _{0}}{2\pi V_h}\,}
e^{-\,8 V_h/\omega_0} \,,
\end{equation}
with $V_h$ the barrier height much smaller than $\omega_0$.
This poses again a curious question;
it appears that
decay of a prepared metastable state localized in the potential
well is never completed.

This simple picture is however valid only when one ignores anharmonic
terms in the tunneling potential, but they must be there
in order to give any realistic tunneling potential.
The most important in the following discussion is effect of
anharmonic terms in the equation for the fluctuation $z(\omega \,, t)$.
Presence of anharmonic terms gives a non-trivial periodicity
in the coefficient function 
$\left(\frac{d^{2} V}{dq^{2} }\right)_{q_{{\rm cl}}^{(0)}}$
for (\ref{fluctuation eq}), assuming a small friction
$\eta \ll \omega_0$.

The homogeneous part of the $z(\omega \,, t)$ solution which
might exhibit the well known parametric resonance \cite{parametric resonance}
is closely related to the behavior for our inhomogeneous 
$z(\omega \,, t)$ solution.
It is then important to check whether 
the relevant parameter in the periodic coefficient function
$\left(\frac{d^{2} V}{dq^{2} }\right)_{q_{{\rm cl}}^{(0)}}$
falls in the instability or the stability band.
As is shown in \cite{my 00-3},
unbounded exponential growth of $\sqrt{I_{11}}$ does not
take place.
On the other hand, the power-law growth is observed in numerical computation.
Moreover, we find that our $z(\omega \,, t)$ solution
belongs to the boundary between stability and instability bands.
At the resonance frequency the enhancement factor due to the boundary
effect is much larger than what one might expect from the harmonic
case (\ref{harmonic z}).
We shall interpret this phenomenon as influenced in a subtle
way by the parametric resonance, although it is not
the parametric resonance itself.

It is best to discuss the resonance enhanced tunneling mechanism
in concrete examples. 
We take the asymmetric wine bottle potential,
which is described in the well and its vicinity region by
\begin{equation}
V(q) \approx \frac{\lambda }{4}(q ^{2} - 2q_{B}q)^{2}
\,.
\end{equation}
The curvature parameters at two extema of $q=0$ and $q = q_B$ are 
\( \:
\omega _{0}^{2} = 2\lambda q_{B}^{2} \,, \hspace{0.5cm} 
\omega _{B}^{2} = \lambda q_{B}^{2} \,,
\: \)
and the barrier height seen from the bottom of the well is
\( \:
V_{h} = \frac{\lambda }{4}q_{B}^{4} = \omega _{0}^{2}q_{B}^{2}/8 \,.
\: \)
The classical $q_{{\rm cl}}^{(0)}$, and the fluctuation
$z(\omega \,, t)$ equations in the Ohmic approximation
are written using rescaled variables, $y = q_{{\rm cl}}^{(0)}/q_{B}$ and
$\tau = \omega_0\,t/2$,
\begin{eqnarray}
&&
y'' + y(y - 1)(y - 2) + \frac{2\eta}{\omega_0}\, y' = 0 \,,
\label{classical ohmic langevin}
 \,,
\\ &&
z'' + \left(\, 4 - 6\,(\,y - \frac{1}{2}\,y^2 \,)\,\right)\,z 
+ \frac{2\eta}{\omega_0}\, z' = -\,\frac{4c(\omega )}{\omega _{0}^{2}}\,
e^{i\,2\omega \tau /\omega _{0}} \,,
\end{eqnarray}
where $' = d/d \tau$.

For a small friction one has approximate forms of solution
in terms of the Jacobi's elliptic function;
\begin{eqnarray}
&&
\hspace*{-1cm}
(y - 1)^{2} = 1 + \sqrt{\epsilon }
 - 2\sqrt{\epsilon }\,{\rm sn}^{2}\left(\,
\sqrt{1 + \sqrt{\epsilon }}\, 
(\,\tau +
\int_{1}^{a}\,\frac{du}{\sqrt{(a^{2} - u^{2})(u^{2} - b^{2})}}\,)
\,, k \right)
 \,, \nonumber \\ &&
\\ &&
\hspace*{-1cm}
p_{{\rm cl}}^{(0)}(t) =
2\sqrt{2 \epsilon V_h}\,{\rm sn}
\left(\,
\sqrt{1 + \sqrt{\epsilon }}\, 
(\,\tau + {\rm const}\,)\,, k \right)
{\rm cn}\left(\,
\sqrt{1 + \sqrt{\epsilon }}\, 
(\,\tau + {\rm const}\,)\,, k \right)
\,, \nonumber \\ &&
\\ &&
\hspace*{0.5cm} 
\tau = \frac{1}{2}\,\omega_0\,t \,,
\hspace{0.5cm}
k = \sqrt{\frac{2\sqrt{\epsilon }}{1 + \sqrt{\epsilon }} }\,, 
\hspace{0.5cm} 
\epsilon = \frac{E_{i}}{V_{h}} 
\,.
\end{eqnarray}
These formulas are valid for 
\( \:
t \ll 1/\eta \,.
\: \)
There are resonances at 
$\tilde{\omega} _{0} \,, 2\tilde{\omega} _{0} \,, \cdots 
n\tilde{\omega} _{0}
\,, \cdots$,
where
\begin{eqnarray}
&&
\tilde{\omega} _{0} = 
\frac{\pi \omega _{0}}{2} \sqrt{1 + \sqrt{\epsilon }}\,
\left( \int_{0}^{1}\,\frac{du}{\sqrt{(1 - u^{2})(1 - k^{2}u^{2})}}\right)
^{-1}
 \,.
\end{eqnarray}

We found so far that these explicit solutions are not very
illuminating, and
numerically integrated the coupled $q_{{\rm cl}}^{(0)}$ 
and $z(\omega \,, t)$ equations.
In the $\omega$ integral (\ref{fluctuation 11}) for the width factor
$I_{11}$ the largest contribution is found to come from
the fundamental resonance at $\omega = \tilde{\omega} _{0}$, 
then contribution from
higher harmonics at $n\,\tilde{\omega} _{0}$ follows.
Thus, phenomenon of a non-linear resonance occurs.
At a time of $O[1/\eta]$ the width
factor $I_{11}$ becomes maximal at its value much larger than in
the harmonic case.
The decrease observed at late times is due to the friction;
we explicitly checked that for the zero friction, 
the fluctuation $|z(\omega \,, t)/c(\omega)|^2$ 
increases in time without a bound, with an averaged time power 
close to 4 at the resonance.
Note that even if the effect of the friction is turned off,
there exists an important environment effect here;
the environment interaction drives the non-linear resonance
oscillation.
We refer to our paper \cite{my 00-3} for detailed numerical
results.

We numerically checked \cite{my 00-3} the behavior of the most important part
given by the product of two competing exponential factors,
\begin{eqnarray}
&&
A = \exp \left[ \,- \,\,\tanh \frac{\beta \omega _{0}}{2}\,
\frac{p_i^2 + \omega_0^2\,q_i^2}{\omega _{0}}
\,\right]
\times
\exp \left[ \,-\,
\frac{(q_B - q_{{\rm cl}}^{(0)}\,)^{2}}{2I_{11}}\,\right]
 \,,
 \label{product factor}
\end{eqnarray}
the one for the initial state and the other for the
kernel factor in the probability rate.
Remarkably, the largest contribution comes, not from the dominant
initial component near the zero point energy, 
rather from the initially suppressed excited component.
The first exponent in (\ref{product factor}) is in proportion to
$E_i$, while the second one goes roughly like $E_i^{-2}$,
hence a maximum may appear somewhere away from the lowest energy
of $E_i = \omega_0/2$.
In the example of $\eta/\omega_0 = 0.0025$ 
the maximum product factor is of order
$10^{-3}$ at $E_i \approx 4 \times \omega_0/2$, 
6 orders of magnitudes larger than
what one expects from the lowest energy state and also
the asymptotic value of order $10^{- 36}$.
A large value of the product factor of order unity 
suggests an interesting possibility of a rapid and violent termination
of the tunneling.

The reason why one observes enhanced tunneling at resonant frequencies of
$\omega = n\,\tilde{\omega}_0$ is as follows.
The semiclassical $q-$motion in the left potential well has
a periodicity $2\pi/\tilde{\omega}_0$ if one neglects the friction.
The environment oscillators act as stochastic fluctuation to this motion
in such a way that the effective potential including one environment
oscillator,
\begin{equation}
V(q) + q\,c(\omega)\,
\left( \,Q_{i}(\omega ) \cos (\omega t) +
\frac{P_{i}(\omega)}{\omega } \sin (\omega t)\,\right)
\,,
\end{equation}
changes the potential barrier periodically.
When the $q-$particle hits against the potential wall,
the potential barrier may become smallest, hence the tunneling rate maximally
enhanced, if one of environment oscillators
has a period exactly equal to that of the classical motion,
$\omega = n\,\tilde{\omega}_0$.
This is precisely the condition of the non-linear resonance
we have been discussing.
The resonance enhanced tunneling thus envisaged seems to have little
connection to the stochastic resonance \cite{stochastic resonance} 
much discussed in the literature.

\vspace{0.5cm} 
In summary, we gave a new mechanism of resonance enhanced
tunneling, using the real-time semiclassical formalism.
Our approach suggests a new time scale of order
$1/\eta $, the inverse friction, for completion of
the first order phase transition.
Evidently much has to be done to apply the idea here to realistic
problems.

\vspace{0.5cm} 
I wish to thank my collaborator, Sh. Matsumoto who
shared most of the results presented here. 

\vspace{1cm} 

\end{document}